\documentclass{amsart}%
\usepackage{amsfonts}
\usepackage{amsmath}
\usepackage{amssymb}
\usepackage{graphicx}%
\theoremstyle{plain}
\newtheorem{acknowledgement}{Acknowledgement}

\newtheorem{definition}{Definition}

\newtheorem{problem}{Problem}
\newtheorem{proposition}{Proposition}
\newtheorem{remark}{Remark}

\numberwithin{equation}{section}
\begin{document}
\title[Is Grover's Algorithm a Quantum Hidden Subgroup Algorithm ?]{Is Grover's Algorithm a Quantum Hidden Subgroup Algorithm ?}
\author{Samuel J. Lomonaco, Jr.}
\address{University of Maryland Baltimore County (UMBC)\\
Baltimore, MD \ 21250 \ \ USA}
\email{Lomonaco@umbc.edu}
\urladdr{http://www.csee.umbc.edu/\symbol{126}lomonaco}
\author{Louis H. Kauffman}
\curraddr{University of Illinois at Chicago\\
Chicago, IL \ 60607-7045 \ \ \ USA}
\email{kauffman@uic.edu}
\urladdr{http://www.math.uic.edu/\symbol{126}kauffman}
\date{March 12, 2006}

\begin{abstract}
The arguments given in this paper suggest that Grover's and Shor's algorithms
are more closely related than one might at first expect. \ Specifically, we
show that Grover's algorithm can be viewed as a quantum algorithm which solves
a non-abelian hidden subgroup problem (HSP). \ But we then go on to show that
the standard non-abelian quantum hidden subgroup (QHS) algorithm can not find
a solution to this particular HSP. \ 

This leaves open the question as to whether or not there is some modification
of the standard non-abelian QHS algorithm which is equivalent to\ Grover's algorithm.

\end{abstract}
\maketitle
\tableofcontents

\section{Introduction}

\bigskip

Is Grover's algorithm a quantum hidden subgroup (QHS) algorithm ? \ 

\bigskip

We do not completely answer this question. \ Instead, we show that Grover's
algorithm is a QHS algorithm in the sense that it can be rephrased as a
quantum algorithm which solves a non-abelian hidden subgroup problem (HSP) on
the symmetric group $\mathbb{S}_{N}$. \ But we then go on to show that the
standard non-abelian QHS algorithm cannot solve the Grover HSP.

\bigskip

This leaves unanswered an intriguing question: \ 

\bigskip

\noindent\textbf{Question.} \textit{Is there an extension or modification of
the standard non-abelian QHS on the symmetric group }$S_{N}$\textit{ which
solves the non-abelian HSP associated with Grover's algorithm?}

\bigskip

It should be mentioned that, because of a result of Zalka \cite{Zalka1}, such
an algorithm, if it exists, could not be asymptotically faster than Grover's
algorithm. \ 

\bigskip

We hope that the results found in this paper will lead to a better
understanding of quantum algorithms.

\bigskip

\section{Definition of the hidden subgroup problem (HSP) and hidden subgroup
algorithms}

\bigskip

What is a hidden subgroup problem ? \ What is a hidden subgroup algorithm ?

\bigskip

\begin{definition}
A map $\varphi:G\longrightarrow S$ from a group $G$ into a set $S$ is said to
have \textbf{hidden subgroup structure} if there exists a subgroup
$K_{\varphi}$ of $G$, called a \textbf{hidden subgroup}, and an injection
$\iota_{\varphi}:G/K_{\varphi}\longrightarrow S$, called a \textbf{hidden
injection}, such that the diagram
\[%
\begin{array}
[c]{ccc}%
G\  & \overset{\varphi}{\longrightarrow} & S\\
\nu\searrow &  & \nearrow\iota_{\varphi}\\
& G/K_{\varphi} &
\end{array}
\]
is commutative\footnote{By saying that this diagram is commutative, we mean
$\varphi=\iota_{\varphi}\circ\nu$. This concept generalizes in an obvious way
to more complicated diagrams.}, where $G/K_{\varphi}$ denotes the collection
of right cosets of $K_{\varphi}$ in $G$, and where $\nu:G\longrightarrow
G/K_{\varphi}$ is the natural surjection of $G$ onto $G/K_{\varphi}$. \ We
refer to the group $G$ as the \textbf{ambient group} and to the set $S$ as the
\textbf{target set}. \ If $K_{\varphi}$ is a normal subgroup of $G$, then
$H_{\varphi}=G/K_{\varphi}$ is a group, called the \textbf{hidden quotient
group}, and $\nu:G\longrightarrow G/K_{\varphi}$ is an epimorphism, called the
\textbf{hidden epimorphism}. \ We will call the above diagram the
\textbf{hidden subgroup structure} of the map $\varphi:G\longrightarrow S$.
\end{definition}

\bigskip

\begin{remark}
The underlying intuition motivating this formal definition is as follows:
\ Given a natural surjection (or epimorphism) $\nu:G\longrightarrow
G/K_{\varphi}$, an "archvillain with malice of forethought" hides the
algebraic structure of $\nu$ by intentionally renaming all the elements of
$G/K_{\varphi}$, and "tossing in for good measure" some extra elements to form
a set $S$ and a map $\varphi:G\longrightarrow S$.
\end{remark}

\bigskip

The hidden subgroup problem can be stated as follows:

\bigskip

\begin{problem}
[\textbf{Hidden Subgroup Problem (HSP)}]Given a map
\[
\varphi:G\longrightarrow S
\]
with hidden subgroup structure, determine a hidden subgroup $K_{\varphi}$ of
$G$. \ An algorithm solving this problem is called a \textbf{hidden subgroup
algorithm}. \ We will call a map with hidden subgroup structure a
\textbf{hidden subgroup problem (HSP).} \ 
\end{problem}

\bigskip

The corresponding quantum form of this HSP is stated as follows:

\bigskip

\begin{problem}
[\textbf{Hidden Subgroup Problem: Quantum Version}]Let
\[
\varphi:G\longrightarrow S
\]
be a map with hidden subgroup structure. \ Construct a quantum implementation
of the map $\varphi$ as follows: \ \medskip

Let $\mathcal{H}_{G}$ and $\mathcal{H}_{S}$ be Hilbert spaces defined
respectively by the orthonormal bases
\[
\left\{  \ \left\vert g\right\rangle \mid g\in G\ \right\}  \text{ and
}\left\{  \ \left\vert s\right\rangle \mid s\in S\ \right\}  \text{ ,}%
\]
and let $s_{0}=\varphi\left(  1\right)  $, where $1$ denotes the identity of
the ambient group $G$. \ Finally, let $U_{\varphi}$ be a unitary
transformation such that
\[%
\begin{array}
[c]{ccc}%
U_{\varphi}:\mathcal{H}_{G}\otimes\mathcal{H}_{S} & \longrightarrow &
\mathcal{H}_{G}\otimes\mathcal{H}_{S}\\
&  & \\
\left\vert g\right\rangle \left\vert s_{0}\right\rangle  & \longmapsto &
\left\vert g\right\rangle \left\vert \varphi\left(  g\right)  \right\rangle
\end{array}
,
\]

Determine the hidden subgroup \ $K_{\varphi}$ with bounded probability of
error by making as few queries as possible of the blackbox $U_{\varphi}$. \ A
quantum algorithm solving this problem is called a \textbf{quantum hidden
subgroup (QHS) algorithm.}
\end{problem}

\bigskip

\section{The generic QHS algorithm \textsc{QRand}}

\bigskip

Let $\varphi:G\longrightarrow S$ be a map from a group $G$ to a set $S$ with
hidden subgroup structure. \ We assume that all representations of $G$ are
equivalent to unitary representations\footnote{This is true for all finite
groups as well as a large class of infinite groups.}. \ Let $\widehat{G}$
denote a \textbf{complete set of distinct irreducible unitary representations}
of $G$. \ Using multiplicative notation for $G$, we let $1$ denote the
\textbf{identity} of $G$, and let $s_{0}$ denote its image in $S$. \ Finally,
let $\widehat{1}$ denote the trivial representation of $G$.

\begin{remark}
If $G$ is abelian, then $\widehat{G}$ becomes the dual group of characters.
\end{remark}

The generic QHS algorithm is given below:\bigskip

\begin{center}
\textbf{Quantum Subroutine} \textsc{QRand}$\left(  \varphi\right)  $ \bigskip
\end{center}

\begin{itemize}
\item[\textbf{Step 0.}] Initialization%
\[
\left\vert \psi_{0}\right\rangle =\left\vert \widehat{1}\right\rangle
\left\vert s_{0}\right\rangle \in\mathcal{H}_{\widehat{G}}\otimes
\mathcal{H}_{S}%
\]

\item[\textbf{Step 1.}] Application of the inverse Fourier transform
$\mathcal{F}_{G}^{-1}$ of $G$ to the left register
\[
\left\vert \psi_{1}\right\rangle =\frac{1}{\sqrt{\left\vert G\right\vert }%
}\sum_{g\in G}\left\vert g\right\rangle \left\vert s_{0}\right\rangle
\in\mathcal{H}_{G}\otimes\mathcal{H}_{S}\text{ \ ,}%
\]
where $\left\vert G\right\vert $ denotes the cardinality of the group
$G$.\bigskip

\item[\textbf{Step 2.}] Application of the unitary transformation $U_{\varphi
}$
\[
\left\vert \psi_{2}\right\rangle =\frac{1}{\sqrt{\left\vert G\right\vert }%
}\sum_{g\in G}\left\vert g\right\rangle \left\vert \varphi\left(  g\right)
\right\rangle \in\mathcal{H}_{G}\otimes\mathcal{H}_{S}\text{ }%
\]
\bigskip

\item[\textbf{Step 3.}] Application of the Fourier transform $\mathcal{F}_{G}$
of $G$ to the left register
\[
\left\vert \psi_{3}\right\rangle =\frac{1}{\left\vert G\right\vert }%
\sum_{\gamma\in\widehat{G}}\left\vert \gamma\right\vert \sum_{g\in
G}Trace\left(  \gamma\left(  g\right)  ^{\dag}\left\vert \gamma\right\rangle
\right)  \left\vert \varphi\left(  g\right)  \right\rangle =\frac
{1}{\left\vert G\right\vert }\sum_{\gamma\in\widehat{G}}\left\vert
\gamma\right\vert Trace\left(  \ \left\vert \gamma\right\rangle \left\vert
\Phi\left(  \gamma^{\dag}\right)  \right\rangle \ \right)  \in\mathcal{H}%
_{\widehat{G}}\otimes\mathcal{H}_{S}\text{ \ ,}%
\]
where $\left\vert \gamma\right\vert $ denotes the degree of the representation
$\gamma$, where $\gamma^{\dag}$ denotes the contragradient representation
(i.e., $\gamma^{\dag}(g)=\gamma\left(  g^{-1}\right)  ^{T}=\overline
{\gamma\left(  g\right)  }^{T}$), where $Trace\left(  \ \gamma^{\dag}\left(
g\right)  \left\vert \gamma\right\rangle \ \right)  =$ $\sum_{i=1}^{\left\vert
\gamma\right\vert }\sum_{j=1}^{\left\vert \gamma\right\vert }\overline
{\gamma\left(  g\right)  }_{ji}\left\vert \gamma_{ij}\right\rangle $, and
where $\left\vert \Phi\left(  \gamma_{ij}^{\dag}\right)  \right\rangle
=\sum_{g\in G}\overline{\gamma}_{ji}\left(  g\right)  \left\vert
\varphi\left(  g\right)  \right\rangle $. \bigskip

\item[\textbf{Step 4.}] Measurement of the left quantum register with respect
to the orthonormal basis
\[
\left\{  \ \left\vert \gamma_{ij}\right\rangle \ :\gamma\in\widehat{G}\text{,
}1\leq i,j\leq\left\vert \gamma\right\vert \right\}  \text{. }%
\]
Thus, with probability
\[
Prob_{\varphi}\left(  \gamma_{ij}\right)  =\frac{\left\vert \gamma\right\vert
^{2}\left\langle \Phi\left(  \gamma_{ij}^{\dag}\right)  |\Phi\left(
\gamma_{ij}^{\dag}\right)  \right\rangle \ }{\left\vert G\right\vert ^{2}%
}\text{ \ ,}%
\]
$\gamma_{ij}$ is the measured result, and the quantum system "collapses" to
the state%
\[
\left\vert \psi_{4}\right\rangle =\frac{\left\vert \gamma_{ij}\right\rangle
\left\vert \Phi\left(  \gamma_{ij}^{\dag}\right)  \right\rangle \ }%
{\sqrt{\ \left\langle \Phi\left(  \gamma_{ij}^{\dag}\right)  |\Phi\left(
\gamma_{ij}^{\dag}\right)  \right\rangle \ }}\in\mathcal{H}_{\widehat{G}%
}\otimes\mathcal{H}_{S}\text{ }%
\]
\bigskip

\item[\textbf{Step 5.}] Output $\gamma_{ij}$ and stop.
\end{itemize}

\bigskip

\section{Pushing HSPs for the generic QHS algorithm \textsc{QRand}}

\bigskip

For certain hidden subgroup problems (HSPs) $\varphi:G\longrightarrow S$, the
corresponding generic QHS algorithm \textsc{QRand} either is not physically
implementable or is too expensive to implement physically. \ For example, the
HSP $\varphi$ is usually not physically implementable if the ambient group is
infinite (e.g., $G$ is the infinite cyclic group $\mathbb{Z}$), and is too
expensive to implement if the ambient group is too large (e.g., $G$ is the
symmetric group $\mathbb{S}_{10^{100}}$). \ In this case, there is a standard
generic way of "tweaking" the HSP to get around this problem, which we will
call \textbf{pushing}.

\bigskip

\begin{definition}
Let $\varphi:G\longrightarrow S$ be a map from a group $G$ to a set $S$. \ A
map $\widetilde{\varphi}:\widetilde{G}\longrightarrow S$ from a group
$\widetilde{G}$ to the set $S$ is said to be a \textbf{push} of $\varphi$,
written
\[
\widetilde{\varphi}=Push\left(  \varphi\right)  \text{ \ ,}%
\]
provided there exists an epimorphism $\mu:G\longrightarrow\widetilde{G}$ from
$G$ onto $\widetilde{G}$, and a transversal $\tau:\widetilde{G}\longrightarrow
G$ of $\mu$ such that $\widetilde{\varphi}=\varphi\circ\tau$.
\end{definition}

\bigskip

If the epimorphism $\mu$ and the transversal $\tau$ are chosen in an
appropriate way, then execution of the generic QHS subroutine with input
$\widetilde{\varphi}=Push\left(  \varphi\right)  $ , i.e., execution of
\[
QRand\left(  \widetilde{\varphi}\right)  \text{ \ ,}%
\]
will with high probability produce an irreducible representation
$\widetilde{\gamma}$ of the group $\widetilde{G}$ which is sufficiently close
to an irreducible representation $\gamma$ of the group $G$. \ If this is the
case, then there is a polynomial time classical algorithm which upon input
$\widetilde{\gamma}$ produces the representation $\gamma$. \ 

\bigskip

Obviously, much more can be said about pushing. \ But unfortunately that would
take us far afield from the objectives of this paper. \ For more information
on pushing, we refer the reader to \cite{Lomonaco7}.

\bigskip

\section{Shor's algorithm}

\bigskip

Shor's factoring algorithm is a classic example of a QHS algorithm created
from the push of an HSP.

\bigskip

Let $N$ be the integer to be factored. Let $\mathbb{Z}$ \ denote the additive
group of integers, and $\mathbb{Z}_{N}^{\times}$ \ denote the monoid of
integers under multiplication modulo $N$ (i.e., the ring of integers modulo
$N$ ignoring addition.) \ 

\bigskip

Shor's algorithm is a QHS algorithm that solves the following HSP
\[%
\begin{array}
[c]{rrc}%
\varphi:\mathbb{Z} & \longrightarrow & \mathbb{Z}_{N}^{\times}\\
m & \longmapsto & a^{m}\operatorname{mod}N
\end{array}
\]
with unknown hidden subgroup structure given by the following commutative
diagram
\[%
\begin{array}
[c]{ccc}%
\mathbb{Z} & \overset{\varphi}{\longrightarrow} & \mathbb{Z}_{N}^{\times}\\
\nu\searrow &  & \nearrow\iota\\
& \mathbb{Z}/P\mathbb{Z} &
\end{array}
\text{ \ ,}%
\]
where $a$ is an integer relatively prime to $N$, where $P$ is the hidden
integer period of the map $\varphi:\mathbb{Z}\longrightarrow\mathbb{Z}%
_{N}^{\times}$, where $P\mathbb{Z}$ is the additive subgroup all integer
multiples of $P$ (i.e., the hidden subgroup), where $\nu:\mathbb{Z}%
\longrightarrow\mathbb{Z}/P\mathbb{Z}$\ is the natural epimorpism of of the
integers onto the quotient group $\mathbb{Z}/P\mathbb{Z}$ (i.e., the hidden
epimorphism), and where $\iota:\mathbb{Z}/P\mathbb{Z\longrightarrow Z}%
_{N}^{\times}$ is the hidden monomorphism.

\bigskip

An obstacle to creating a physically implementable algorithm for this HSP is
that the domain $\mathbb{Z}$ of $\varphi$ is infinite. \ As observed by Shor,
a way to work around this difficulty is to push the HSP.

\bigskip

In particular, as illustrated by the following commutative diagram%
\[%
\begin{array}
[c]{ccl}%
\mathbb{Z\qquad} & \overset{\varphi}{\longrightarrow} & \qquad\mathbb{Z}%
_{N}^{\times}\\
\mu\searrow\nwarrow\tau &  & \nearrow\varphi=Push\left(  \varphi\right)
=\varphi\circ\tau\\
& \mathbb{Z}_{Q} &
\end{array}
\text{ \ \ ,}%
\]
a push $\widetilde{\varphi}=Push\left(  \varphi\right)  $ is constructed by
selecting the epimorphism $\mu:\mathbb{Z\longrightarrow Z}_{Q}$ of
$\mathbb{Z}$ onto the finite cyclic group $\mathbb{Z}_{Q}$ of order $Q$, where
the integer $Q$ is the unique power of $2$ such that $N^{2}\leq Q<2N^{2}$, and
choosing the transversal\footnote{A \textbf{transversal} for an epimorphism
$\alpha_{\varphi}:\mathbb{Z\longrightarrow Z}_{Q}$ is an injection
$\tau_{\varphi}:\mathbb{Z_{Q}\longrightarrow Z}$ such that $\alpha_{\varphi
}\circ\tau_{\varphi}$ is the identity map on $\mathbb{Z}_{Q}$, i.e., a map
that takes each element of $\mathbb{Z}_{Q}$ onto a coset representative of the
element in $\mathbb{Z}$ .}
\[%
\begin{array}
[c]{rrc}%
\tau:\mathbb{Z}_{Q} & \longrightarrow & \mathbb{Z}\\
m\operatorname{mod}Q & \longmapsto & m
\end{array}
\text{ \ ,}%
\]
where $0\leq m<Q$. \ \textit{This push} $\widetilde{\varphi}=Push\left(
\varphi\right)  $ \textit{is called} \textbf{Shor's oracle}.

\bigskip

Shor's algorithm consists in first executing the quantum subroutine
\textsc{QRand}$\left(  \widetilde{\varphi}\right)  $, thereby producing a
random character
\[
\gamma_{y/Q}:m\operatorname{mod}Q\mapsto\frac{my}{Q}\operatorname{mod}1
\]
of the finite cyclic group $\mathbb{Z}_{Q}$. \ The transversal $\tau$ used in
pushing has been engineered to assure that the character $\gamma_{y/Q}$ is
sufficiently close to a character
\[
\gamma_{d/P}:k\operatorname{mod}P\mapsto\frac{kd}{P}\operatorname{mod}1
\]
of the hidden quotient group $\mathbb{Z}/P\mathbb{Z}=\mathbb{Z}_{P}$. \ In
this case "sufficiently close" means that
\[
\left\vert \frac{y}{Q}-\frac{d}{P}\right\vert \leq\frac{1}{2P^{2}}\text{ \ ,}%
\]
which that $d/P$ is a continued fraction convergent of $y/Q$, and thus can be
found found by the classical polynomial time continued fraction algorithm.

\bigskip

\section{Description of Grover's algorithm}

\bigskip

Now let us turn to Grover's algorithm. \ We begin with a brief description.

\bigskip

Consider an unstructured database of $N=2^{n}$ records labeled without
repetitions with the labels
\[
0,1,2,\ldots,N-1\text{. }%
\]
We are given the oracle $f:\left\{  0,1\right\}  ^{n}\longrightarrow\left\{
0,1\right\}  $, where%
\[
f(x)=\left\{
\begin{array}
[c]{cll}%
1 & \text{if }j=j_{0} & \text{(\textquotedblleft Yes\textquotedblright)}\\
&  & \\
0 & \text{otherwise} & \text{(\textquotedblleft No\textquotedblright) \ \ \ ,}%
\end{array}
\right.
\]
called \textbf{Grover's oracle}, and asked to solve the following search problem:

\bigskip

\noindent\textbf{Search Problem for an Unstructured Database.} \ \textit{Find
the unknown record labeled as }$j_{0}$\textit{\ with the minimum amount of
computational work, i.e., with the minimum number of queries of the oracle
}$f$, and with bounded probability of error\textit{.}

\bigskip\bigskip

Let $\mathcal{H}$ be the Hilbert space with orthonormal basis
\[
\left\vert 0\right\rangle ,\left\vert 1\right\rangle ,\left\vert
2\right\rangle ,\ldots,\left\vert N-1\right\rangle \text{ ,}%
\]
where $N=2^{n}$. \ Then Grover's oracle is essentially given as the unitary
transformation
\[%
\begin{array}
[c]{cccl}%
I_{\left\vert j_{0}\right\rangle }: & \mathcal{H} & \longrightarrow &
\mathcal{H}\\
\multicolumn{1}{r}{} & \multicolumn{1}{r}{\left\vert j\right\rangle } &
\multicolumn{1}{r}{\longmapsto} & \left(  -1\right)  ^{f(j)}\left\vert
j\right\rangle
\end{array}
\]
where
\[
I_{\left\vert j_{0}\right\rangle }=I-2\left\vert j_{0}\right\rangle
\left\langle j_{0}\right\vert
\]
is inversion in the hyperplane orthogonal to $\left\vert j_{0}\right\rangle $.

\bigskip

Let $H$ denote the Hadamard transform on the Hilbert space $\mathcal{H}$.
\ Then Grover's algorithm is given as:

\bigskip

\begin{center}
\fbox{%
\begin{tabular}
[c]{ll}\hline\hline
& \hspace{0.75in}\textbf{Grover's Algorithm}\\\hline\hline
& \\
$%
\begin{array}
[c]{r}%
\fbox{$\mathbb{STEP}$ 0.}\\
\bigskip\\
\bigskip
\end{array}
$ & $%
\begin{array}
[c]{l}%
\text{(Initialization)}\\
\qquad\left\vert \psi\right\rangle \longleftarrow H\left\vert 0\right\rangle
=\frac{1}{\sqrt{N}}%
{\displaystyle\sum\limits_{j=0}^{N-1}}
\left\vert j\right\rangle \\
\qquad k\quad\longleftarrow0
\end{array}
$\\
& \\
$%
\begin{array}
[c]{r}%
\fbox{$\mathbb{STEP}$ 1.}\\
\bigskip\\
\bigskip
\end{array}
$ & $%
\begin{array}
[c]{r}%
\text{Loop until }k=\underset{}{\left\lfloor \frac{\pi}{4\sin^{-1}\left(
1/\sqrt{N}\right)  }\right\rfloor }\approx\left\lfloor \frac{\pi}{4}\sqrt
{N}\right\rfloor \\
\multicolumn{1}{l}{\qquad\left\vert \psi\right\rangle \longleftarrow
\underset{}{Q}\left\vert \psi\right\rangle =-HI_{\left\vert 0\right\rangle
}HI_{\left\vert j_{0}\right\rangle }\left\vert \psi\right\rangle }\\
\multicolumn{1}{l}{\qquad k\quad\longleftarrow k+1}%
\end{array}
$\\
& \\
$%
\begin{array}
[c]{r}%
\fbox{$\mathbb{STEP}$ 2.}\\
\bigskip
\end{array}
$ & $%
\begin{array}
[c]{l}%
\text{Measure }\left\vert \psi\right\rangle \text{ with respect to the
standard basis}\\
\left\vert 0\right\rangle ,\left\vert 1\right\rangle ,\ \ldots\ ,\left\vert
N-1\right\rangle \text{ to obtain the unknown }\\
\text{state }\left\vert j_{0}\right\rangle \text{ with probability }%
\geq1-\frac{1}{N}\text{.}%
\end{array}
$%
\end{tabular}
}
\end{center}

\bigskip

\section{The symmetry hidden within Grover's algorithm}

\bigskip

But where is the \textit{hidden symmetry} in Grover's algorithm ?

\bigskip

Let $\mathbb{S}_{N}$ be the symmetric group on the symbols
\[
0,1,2,3,\ldots,N-1\text{ .}%
\]
Then Grover's algorithm is invariant under the \textbf{hidden subgroup}
\[
Stab_{j_{0}}=\left\{  g\in\mathbb{S}_{N}:g\left(  j_{0}\right)  =j_{0}%
\right\}  \subset\mathbb{S}_{N}\text{ ,}%
\]
called the \textbf{stabilizer subgroup} for $j_{0}$, i.e., Grover's algorithm
is invariant under the group action%
\[%
\begin{array}
[c]{ccc}%
Stab_{j_{0}}\times\mathcal{H} & \longrightarrow & \mathcal{H}\\
&  & \\
\left(  g,\sum_{j=0}^{N-1}a_{j}\left\vert j\right\rangle \right)  &
\longmapsto & \sum_{j=0}^{N-1}a_{j}\left\vert g(j)\right\rangle
\end{array}
\]
Moreover, if the hidden subgroup $Stab_{j_{0}}$ is known, then so is the
integer $j_{0}$, and vice versa. \ 

\bigskip

Thus, Grover's algorithm is an algorithm that solves the following hidden
subgroup problem, which we will henceforth refer to as \textbf{Grover's hidden
subgroup problem}:

\bigskip\bigskip

\noindent\textsc{Grover}'\textsc{s Hidden Subgroup Problem.} \ \textit{Given a
map}%
\[
\mathbb{S}_{N}\overset{\varphi}{\longrightarrow}S
\]
\textit{from the the symmetric group }$\mathbb{S}_{N}$\textit{ into a target
set }$S=\left\{  0,1,2,\ldots,N-1\right\}  $\textit{ with hidden subgroup
structure given by the commutative diagram }%
\[%
\begin{array}
[c]{ccc}%
\mathbb{S}_{N} & \overset{\varphi}{\longrightarrow} & S\\
\nu_{j_{0}}\searrow &  & \nearrow\iota\\
& \mathbb{S}_{N}/Stab_{j_{0}} &
\end{array}
\text{ \ ,}%
\]
\textit{where }$\nu_{j_{0}}:S_{N}\longrightarrow S/Stab_{j_{0}}$\textit{ is
the natural surjection of }$S_{N}$\textit{ onto the coset space }%
$S/Stab_{j_{0}}$\textit{, and where }%
\[%
\begin{array}
[c]{rrr}%
\iota:\mathbb{S}_{N} & \overset{\varphi}{\longrightarrow} & S\\
\left(  j\ j_{0}\right)  Stab_{j_{0}} & \longmapsto & j
\end{array}
\]
\textit{is the \textbf{unknown relabeling} (bijection) of the coset space
}$S_{N}/Stab_{j_{0}}$\textit{ onto the set }$S$\textit{. \ Find the hidden
subgroup }$Stab_{j_{0}}$\textit{ with bounded probability of error.}

\bigskip\bigskip

Let $\left(  ij\right)  \in\mathbb{S}_{N}$ denote the permutation that
interchanges integers $i$ and $j$, and leaves all other integers fixed.
\ Thus, $\left(  ij\right)  $ is a transposition if $i\neq j$, and the
identity permutation $1$ if $i=j$. \ 

\bigskip

\begin{proposition}
The set%
\[
\left\{  \left(  0j_{0}\right)  ,\left(  1j_{0}\right)  ,\left(
2j_{0}\right)  ,\ldots,\left(  \left(  N-1\right)  j_{0}\right)  \right\}
\]
is a complete set of distinct coset representatives for the hidden subgroup
$Stab_{j_{0}}$ of $\mathbb{S}_{N}$, i.e., the coset space $\mathbb{S}%
_{N}/Stab_{j_{0}}$ is given by the following complete set of mutually distinct
cosets.\bigskip%
\[
\mathbb{S}_{N}/Stab_{j_{0}}=\left\{  \left(  0j_{0}\right)  Stab_{j_{0}%
},\left(  1j_{0}\right)  Stab_{j_{0}},\left(  2j_{0}\right)  Stab_{j_{0}%
},\ldots,\left(  \left(  N-1\right)  j_{0}\right)  Stab_{j_{0}}\right\}
\]

\end{proposition}

\begin{proof}
Since\bigskip%
\[
\left(  kj_{0}\right)  Stab_{j_{0}}=\left(  \ell j_{0}\right)  Stab_{j_{0}%
}\Longleftrightarrow\left(  \ell j_{0}\right)  ^{-1}\left(  kj_{0}\right)  \in
Stab_{j_{0}}\Longleftrightarrow k=l\text{ ,}%
\]
\bigskip it follows that
\[
\left(  0j_{0}\right)  Stab_{j_{0}},\left(  1j_{0}\right)  Stab_{j_{0}%
},\left(  2j_{0}\right)  Stab_{j_{0}},\ldots,\left(  \left(  N-1\right)
j_{0}\right)  Stab_{j_{0}}%
\]
are mutually distinct cosets of $Stab_{j_{0}}$ in $\mathbb{S}_{N}$. \ It now
follows from Lagrange's theorem that the above collection of mutually distinct
cosets is complete.
\end{proof}

\bigskip

\section{A comparison of Grover's and Shor's algorithms}

\bigskip

Now let us compare Shor's algorithm with Grover's.

\bigskip

Let $S$ be the set of integers%
\[
S=\left\{  0,1,2,\ldots,N-1\right\}  \text{ \ ,}%
\]
where $N=2^{n}$, and let $j_{0}\in S$ denote the unknown label to be found by
Grover's algorithm.

\bigskip

Shor's algorithm solves the HSP $\varphi:\mathbb{Z}\longrightarrow
\mathbb{Z}_{N}^{\times}$ with hidden subgroup structure
\[%
\begin{array}
[c]{ccc}%
\mathbb{Z} & \overset{\varphi}{\longrightarrow} & \mathbb{Z}_{N}^{\times}\\
\nu\searrow &  & \nearrow\iota\\
& \mathbb{Z}/P\mathbb{Z} &
\end{array}
\text{ \ ,}%
\]
where $\mathbb{Z}_{N}^{\times}$ can be thought of as the result of the unknown
("malicious") relabeling $\ $%
\[
\iota:k+PZ\longmapsto a^{k}\operatorname{mod}N
\]
of $\mathbb{Z}/P\mathbb{Z}$.

\bigskip

In like manner, Grover's algorithm solves an HSP, namely, the HSP
$\varphi:\mathbb{S}_{N}\longrightarrow S$ with hidden subgroup structure
\[%
\begin{array}
[c]{ccc}%
\mathbb{S}_{N} & \overset{\varphi}{\longrightarrow} & S\\
\nu\searrow &  & \nearrow\iota\\
& \mathbb{S}_{N}/Stab_{j0} &
\end{array}
\text{ \ ,}%
\]
where $S=\left\{  0,1,2,\ldots,N-1\right\}  $ denotes the set resulting from
the unknown ("malicious") relabeling (bijection)%
\[
\iota:\left(  j\ j_{0}\right)  Stab_{j_{0}}\longmapsto j
\]
of $\mathbb{S}_{N}/Stab_{j0}$.

\bigskip

For Shor's algorithm, Shor's oracle $\widetilde{\varphi}:\mathbb{Z}%
_{Q}\longrightarrow\mathbb{Z}_{N}^{\times}$ is created by pushing the HSP
$\varphi:\mathbb{Z}\longrightarrow\mathbb{Z}_{N}^{\times}$ using
\[%
\begin{array}
[c]{ccc}%
\mathbb{Z} & \overset{\varphi}{\longrightarrow} & \mathbb{Z}_{N}^{\times}\\
\mu\searrow\nwarrow\tau &  & \nearrow\widetilde{\varphi}\\
& \mathbb{Z}/Q\mathbb{Z} &
\end{array}
\text{ \ ,}%
\]
thereby producing $\widetilde{\varphi}=Push(\varphi)=\varphi\circ\tau$ with
the transversal $\tau:k\operatorname{mod}Q\longmapsto k$.

\bigskip

In like manner, for Grover's algorithm, Grover's oracle can be created by
pushing the HSP $\varphi:\mathbb{S}_{N}\longrightarrow S$ using
\[%
\begin{array}
[c]{ccc}%
\mathbb{S}_{N} & \overset{\varphi}{\longrightarrow} & S\\
\mu\searrow\nwarrow\tau &  & \nearrow\widetilde{\varphi}\\
& \mathbb{S}_{N}/Stab_{0} &
\end{array}
\text{ \ ,}%
\]
thereby producing $\widetilde{\varphi}=Push(\varphi)=\varphi\circ\tau$ with
the transversal $\tau:\left(  0\ j\right)  Stab_{0}\longmapsto\mathbb{S}_{N}$
of the natural surjection $\mu$.

\bigskip

Although it is not immediately apparent, the resulting push $\widetilde
{\varphi}$ (for $j_{0}\neq0$) is actually Grover's oracle relabelled by the
injection $\iota:S_{N}/Stab_{j_{0}}\longrightarrow S$. \ For $\widetilde
{\varphi}=\varphi\circ\tau=\left(  \iota\circ\nu\right)  \circ\tau=\iota
\circ\left(  \nu\circ\tau\right)  $ and%
\[
\left(  \nu\circ\tau\right)  \left[  \left(  0\ j\right)  Stab_{0}\right]
=\left\{
\begin{array}
[c]{ll}%
\left(  0\ j_{0}\right)  Stab_{j_{0}} & \text{if \ }j=j_{0}\\
& \\
Stab_{j_{0}} & \text{otherwise}%
\end{array}
\right.
\]
which is informationally the same as Grover's oracle
\[
f\left(  j\right)  =\left\{
\begin{array}
[c]{ll}%
1 & \text{if \ }j=j_{0}\\
& \\
0 & \text{otherwise}%
\end{array}
\right.
\]

\bigskip

Hence, we can conclude that Grover's algorithm is an quantum algorithm very
much like Shor's algorithm, in that it is a quantum algorithm that solves the
Grover hidden subgroup problem.\bigskip

\section{However}

\bigskip

However, ... this appears to be where the similarity between these two
algorithms ends. \ For, the standard non-abelian QHS algorithm on
$\mathbb{S}_{N}$ for the HSP $\varphi$ (or $\widetilde{\varphi}$) can not find
the hidden subgroup $Stab_{j_{0}}$ for each of the following two reasons:

\bigskip

\begin{itemize}
\item Since the subgroups $Stab_{j}$ are not normal subgroups of
$\mathbb{S}_{N}$, it follows from the work of Hallgren et al \cite{Hallgren1}
that the standard non-abelian hidden subgroup algorithm will find the largest
normal subgroup of $\mathbb{S}_{N}$ lying in $Stab_{j}$. \ But unfortunately,
the largest normal subgroup of $\mathbb{S}_{N}$ lying in $Stab_{j}$. is the
trivial subgroup of $\mathbb{S}_{N}$.\bigskip

\item The subgroups $Stab_{0}$, $Stab_{1}$, ... , $Stab_{N-1}$ are mutually
conjugate subgroups of $\mathbb{S}_{N}$.
\end{itemize}

\bigskip

We should also mention that this hidden subgroup approach can not possibly
lead to a quantum algorithm that is faster than Grover's.. \ For
Zalka\cite{Zalka1} has shown that Grover's algorithm is asymptotically optimal.

\bigskip%
\[%
\begin{tabular}
[c]{|c|c|}\hline
\multicolumn{2}{|c|}{\textbf{A Comparison of Two Quantum Algorithms}}\\\hline
\textbf{Shor's Algorithm} & G\textbf{rover's Algorithm}\\\hline
\multicolumn{2}{|c|}{\textbf{Similarities\quad}}\\\hline%
\begin{tabular}
[c]{l}%
Shor's algorithm solves\\
an HSP, namely:\\
\multicolumn{1}{c}{%
\begin{tabular}
[c]{l}%
\begin{tabular}
[c]{lll}%
$\mathbb{Z}$ & $\overset{\varphi}{\longrightarrow}$ & $\mathbb{Z}_{N}^{\times
}$\\
$\nu\searrow$ &  & $\nearrow\iota$\\
& $\mathbb{Z}/P\mathbb{Z}$ &
\end{tabular}
\end{tabular}
}\\
\end{tabular}
&
\begin{tabular}
[c]{l}%
Grover's algorithm solves\\
an HSP, namely:\\
\multicolumn{1}{c}{%
\begin{tabular}
[c]{l}%
$%
\begin{array}
[c]{rrr}%
\mathbb{S}_{N} & \overset{\varphi}{\longrightarrow} & S\\
\nu\searrow &  & \nearrow\iota\\
& \mathbb{S}_{N}/Stab_{j_{0}} &
\end{array}
$%
\end{tabular}
}\\
\end{tabular}
\\\hline%
\begin{tabular}
[c]{l}%
Pushing $\varphi$ using\\
\multicolumn{1}{c}{%
\begin{tabular}
[c]{l}%
$%
\begin{array}
[c]{rrr}%
\mathbb{Z} & \overset{\varphi}{\longrightarrow} & \mathbb{Z}_{N}^{\times}\\
\mu\searrow\nwarrow\tau &  & \nearrow\widetilde{\varphi}\\
& \mathbb{Z}/Q\mathbb{Z} &
\end{array}
$%
\end{tabular}
}\\
\multicolumn{1}{c}{produces $\widetilde{\varphi}=Push(\varphi)=\varphi
\circ\tau$}\\
which is Shor's oracle
\end{tabular}
&
\begin{tabular}
[c]{l}%
Pushing $\varphi$ using\\%
\begin{tabular}
[c]{l}%
$%
\begin{array}
[c]{rrr}%
\mathbb{S}_{N} & \overset{\varphi}{\longrightarrow} & S\\
\mu\searrow\nwarrow\tau &  & \nearrow\widetilde{\varphi}\\
& \mathbb{S}_{N}/Stab_{0} &
\end{array}
$%
\end{tabular}
\\
produces $\widetilde{\varphi}=Push(\varphi)=\varphi\circ\tau$\\
which is Grover's oracle ($j_{0}\neq0$)
\end{tabular}
\\\hline
\multicolumn{2}{|c|}{\textbf{Differences}}\\\hline%
\begin{tabular}
[c]{l}%
Repeated calling of the quantum\\
subroutine \textsc{QRand}$\left(  \widetilde{\varphi}\right)  $ provides\\
enough information to solve the\\
HSP $\varphi$%
\end{tabular}
&
\begin{tabular}
[c]{l}%
Repeated calling of the quantum\\
subroutine \textsc{QRand}$\left(  \widetilde{\varphi}\right)  $ provides\\
no information whatsoever about\\
the HSP $\varphi$%
\end{tabular}
\\\hline
\end{tabular}
\]

\bigskip

\section{Conclusions and Open Questions}

\bigskip

The arguments made in this paper suggest that Grover's and Shor's algorithms
are more closely related quantum algorithms than one might at first expect.
\ Although the standard non-abelian QHS algorithm on $\mathbb{S}_{N}$ can not
solve the Grover hidden subgroup problem, there still remains an intriguing question:

\bigskip

\noindent\textbf{Question.} \ \textit{Is there some modification or extension
of the stantard non-abelian QHS algorithm on the symmetric group }%
$\mathbb{S}_{N}$\textit{ that actually solves Grover's hidden subgroup
problem?}

\bigskip

An answer to the above question could lead to a greater insight into how to
create new quantum algorithms.\bigskip

\bigskip

The methods of this paper can also be applied to Grover's algorithm for
multiple marked label search. \ But can they also be applied to other
extensions of Grover's algorithm such as those found in \cite{Biham1},
\cite{Biham2}?

\bigskip

\begin{acknowledgement}
This work is partially supported by the Defense Advanced Research Projects
Agency (DARPA) and Air Forche Research Laboratory, Air Force Materiel Command,
USAF, under agreement number F30602-01-2-0522. The U.S. Government is
authorized to reproduce and distribute reprints for Governmental purposes
notwithstanding any copyright annotation thereon. \ This work also partially
supported by the Institute for Scientific Interchange (ISI), Torino, the
National Institute of Standards and Technology (NIST), the Mathematical
Sciences Research Institute (MSRI), the Isaac Newton Institute for
Mathematical Sciences, and the L-O-O-P fund.
\end{acknowledgement}

\end{document}